\documentclass[apjl]{emulateapj}
\usepackage{apjfonts}

\shorttitle{1RXS~J1609-2105~b: confirmation of common proper motion}
\shortauthors{Lafreni\`{e}re et al.}

\newcommand{\msun}{\ensuremath{M_{\odot}}}
\newcommand{\mjup}{\ensuremath{M_{\rm Jup}}}

\newcommand\planethost{1RXS~J160929.1-210524}
\newcommand\primary{1RXS~J1609-2105}
\newcommand\companion{1RXS~J1609-2105~b}

\begin{document}

\title{The Directly Imaged Planet around the Young Solar Analog \planethost: Confirmation of Common Proper Motion, Temperature and Mass}

\author{David Lafreni\`ere\altaffilmark{1,2}, Ray Jayawardhana\altaffilmark{2}, Marten H. van Kerkwijk\altaffilmark{2}}
\altaffiltext{1}{D\'epartement de physique, Universit\'e de Montr\'eal, C.P. 6128 Succ. Centre-Ville, Montr\'eal, QC, H3C 3J7, Canada}
\altaffiltext{2}{Department of Astronomy and Astrophysics, University of Toronto, 50 St. George Street, Toronto, ON, M5S 3H4, Canada}
\email{david@astro.umontreal.ca}

\begin{abstract}

Giant planets are usually thought to form within a few tens of AU of their host stars, and hence it came as a surprise when we found what
appeared to be a planetary mass ($\sim$ 0.008 $\msun$) companion around the 5 Myr-old solar mass star \planethost\ in the Upper Scorpius association. At the time, we took the object's membership in Upper Scorpius ---established from near-infrared, $H$- and $K$-band spectroscopy--- and its proximity (2.2\arcsec, or 330~AU) to the primary as strong evidence for companionship, but could not verify their common proper motion.  Here, we present follow-up astrometric measurements that confirm that the companion is indeed co-moving with the primary star, which we interpret as evidence that it is a truly bound planetary mass companion. We also present new $J$-band spectroscopy and 3.0-3.8~$\mu$m photometry of the companion. Based on a comparison with model spectra, these new measurements are consistent with the previous estimate of the companion effective temperature of $1800\pm200$~K. We present a new estimate of the companion mass based on evolution models and the calculated bolometric luminosity of the companion; we obtain a value of $0.008_{-0.002}^{+0.003}$~\msun, again consistent with our previous result. Finally, we present angular differential imaging observations of the system allowing us to rule out additional planets in the system more massive than 1~\mjup, 2~\mjup\ and 8~\mjup\ at projected separations larger than 3\arcsec\ ($\sim$440~AU), 0.7\arcsec\ ($\sim$100~AU) and 0.35\arcsec\ ($\sim$50~AU), respectively. This companion is the least massive known to date at such a large orbital distance; it shows that objects in the planetary mass range exist at orbital separations of several hundred AU, posing a serious challenge for current formation models.

\end{abstract} 

\section{Introduction}

The detection and characterization of low-mass substellar companions on large orbits (over several tens of AU) around stars is of great importance for our understanding of planet, brown dwarf and star formation, as well as for our understanding of the dynamical evolution of such companions in multiple systems or in circumstellar disks. Currently most formation models of planets or low-mass brown dwarf companions -- core accretion \citep[e.g.][]{pollack96}, gravitational instability \citep[e.g.][]{boss97} and fragmentation of a pre-stellar core \citep[e.g.][]{padoan02, bate03} -- cannot easily explain the existence of such companions at large orbital separations. A good determination of the semi-major axis, mass and mass ratio distributions of far out giant planets will help us understand these mechanisms in more detail and disentangle the role played by each one. 

Over the past decade, observations of young stars have led to the discovery of several companions with separations of several tens to hundreds of AU and masses only slightly above the $\sim$13~\mjup\  deuterium burning limit: TWA~5~B \citep{lowrance99}, AB~Pic~B \citep{chauvin05b}, GQ~Lub~B \citep{neuhauser05}, DH~Tau~B \citep{itoh05}, CHXR~73~B \citep{luhman06}, LP~261-75~B \citep{reid06}, HN~Peg~B \citep{luhman07}, and CT~Cha~B \citep{schmidt08}. For higher mass brown dwarf companions, the list of known companions with similar orbital separations is even longer, the reader is referred to \citet{zuckerman09} for a recent compilation. More recently, the mass limit of known substellar companions on very wide orbits has been pushed down below the deuterium burning limit ---i.e. in the planetary range--- by a rapid succession of discoveries, an $\sim$8~\mjup\ planetary mass object around the young solar analog \planethost\ (\citealp{lafreniere08b}; hereafter called Paper~I), three $\sim$7--10~\mjup\ planets around the A-type star HR~8799 \citep{marois08}, a ~$\sim$1~\mjup\ planet orbiting inside the dust belt of the A-type star Fomalhaut \citep{kalas08}, and an $\sim$8~\mjup\ planet candidate within the circumstellar disk of the A-type star $\beta$~Pictoris \citep{lagrange09a}. The newly-found companion to the Sun-like star GJ~758 \citep{thalmann09}, a 10--40~\mjup\ object orbiting at a separation of 29~AU, could add to the list of planetary mass companions if its true age turns out to be at the young end of current estimates. A few planetary mass objects have also been found in large orbits around brown dwarf primaries: 2MASS~1207b \citep{chauvin05a}, 2MASS~0441b \citep{todorov10} and possibly UScoCTIO~108~b \citep{bejar08}. Beside actual planet detections, observations of asymmetries or ring-like structures in some debris disks offer indirect evidence for the presence of planets on large orbits around stars. Good examples include $\beta$~Pic, HR~4796, $\varepsilon$~Eri, Vega, HD~141569, $\eta$~Corvi, Fomalhaut, and AB~Aur (\citealp{holland98, greaves98, weinberger99, schneider99, kalas05, wyatt05, oppenheimer08}; see also \citealt{wyatt08}). The peculiar disk morphologies observed in those systems could arise from gravitational perturbations by unseen planets. Modeling indicates that Neptune to Jupiter mass planets at several tens of AUs are typically needed to reproduce the observations. Although the current picture is still far from complete, all of these results offer a very interesting first glimpse of the full extent of orbital separations and masses of giant planets and low-mass brown dwarf companions around stars.

The planetary mass candidate companion to the young solar mass star \planethost, which is a member of the 5 Myr-old Upper Scorpius association \citep{preibisch02} is the subject of this paper. The candidate companion, lying only 2.2\arcsec ($\sim$330~AU projected) from the primary star, appears to be a young planetary mass ($0.008$~\msun) object in the Upper~Sco association, based on near-infrared imaging ($JHK$) and spectroscopy ($HK$). In particular, the triangular shape of its $H$-band spectrum clearly indicates low surface gravity, and the overall spectrum is in excellent agreement with the spectra of known low-mass brown dwarfs in the same association as well as with model spectra with $T_{\rm eff}\sim1800$~K and $\log{g}\sim4$, consistent with the inferred age and mass. While at the time of discovery, we had no proper motion measurements and hence could not confirm it was gravitationally bound, the proximity and youth of the candidate companion offered very compelling evidence that the two form a bound pair. In this paper, we present follow-up astrometric observations that confirm the common proper motion of the candidate companion with the primary star. In addition, to improve the characterization of this system, we present new photometric observations at 3.05~$\mu$m and 3.78~$\mu$m ($L^\prime$), new spectroscopic observations in the $J$-band, and new high-contrast angular differential imaging (ADI, \citealp{marois06}) observations.

\section{Observations and data reduction} \label{sect:obs}

All observations used for this paper\footnote{In addition to the observations described in this paper, we also obtained exploratory VLA observations at 21 and 6\,cm as the NVSS catalog lists a radio source with a brightness of 5\,mJy near the source. However the radio source is not associated; at 6\,cm, the 0.7\,mJy point source is at $\alpha_{\rm J2000}=16:09:29.6772\pm0.0011$ and $\delta_{\rm J2000}=-21:04:49.591\pm0.027$.} were obtained with the Gemini North telescope and the ALTAIR adaptive optics system \citep{herriot00}. The target star itself was used for wave front sensing and the field lens of ALTAIR was used to reduce off-axis Strehl degradation due to anisoplanatism. 

\subsection{Imaging}

All imaging observations were obtained with the NIRI camera \citep{hodapp03} in $f/32$ mode. Used with the AO system and the field lens, this yields a pixel scale of 21.4~mas and a field of view of $21.9\arcsec \times 21.9\arcsec$ for the full detector array.

The initial $JHK$ imaging observations of \primary\ were obtained in 2008 April and June and are detailed in Paper~I. For astrometric follow-up of the candidate companion, the star was observed again on 2009 April 6 and 2009 July 1. These observations were made in the $K^\prime$ filter and used a 5-point dither pattern corresponding to the center and four corners of a square of side 10\arcsec. At each dither position, we obtained one co-addition of thirty 0.3~s integrations in fast, high read-noise mode, and one 10~s integration in slow, low read-noise mode. At each position this provides an unsaturated image of the target star and a much deeper image of the field that can be spatially registered and scaled in flux without ambiguity.

These imaging data were reduced using custom {\em IDL} routines. A sky frame was constructed by taking the median of the images at all dither positions after masking out the regions dominated by the target's signal. After subtraction of this sky frame, the images were divided by a normalized flat-field. Then isolated bad pixels were replaced by the interpolated value of a third-order polynomial surface fit to the good pixels in a $7\times7$ pixels box while clustered bad pixels were simply masked out. Next the images were distortion corrected using the distortion solution provided by the Gemini staff.\footnote{The distortion correction used is centro-symmetric and is given by $r=r^\prime+1.32\times10^{-5}r^{\prime 2}$, where $r$ and $r^\prime$ are respectively the distortion-corrected and distorted radial pixel distances from the array center.} Finally, the long-exposure images were properly scaled in intensity and merged with their corresponding short-exposure images.

Follow-up observations of the system were also obtained at 3.05~$\mu$m (H$_2$O~ice filter, 5\% bandpass) and 3.78~$\mu$m ($L^\prime$ filter) on 2009 May 2 and 2009 June 8, respectively, to constrain the spectral energy distribution of the candidate companion further. For 3.05~$\mu$m, 20 co-additions of ten 5~s integrations were obtained over 20 dither positions covering a square of side 8\arcsec. To enable a shorter integration time and avoid saturation for 3.78~$\mu$m, only the central $768\times768$ pixels of the array was read and 30 co-additions of three hundred 0.2~s integrations were obtained over 30 dither positions covering a square of side 6\arcsec.

For the long wavelength imaging, the data reduction involved first removing the sky background by subtracting, for each image, the mean of the two images acquired closest in time for which the star is not located in the vicinity of the candidate companion in the image considered. The images were then divided by a normalized flat field image created by taking the median of all images after masking out the star in each one and normalizing them to a median value of one. A $\sim 121 \times 121$ pixels median-filtered image was subsequently subtracted from each image to remove any low spatial frequency residual background signal, which was occasionally present; to avoid any potential bias in doing this, the median filter was computed after masking out a $9\times 9$ pixel box over the star and companion. All images were finally aligned and combined using a 3$\sigma$ clipped average.

An angular differential imaging (ADI, \citealt{marois06}) sequence of observations was obtained on 2009 April 6 to search for additional, closer and/or fainter companions in the system. These observations were done in the $K^\prime$ filter and only the central $512\times512$ pixel region of the detector was read. The main ADI sequence consisted of 320 exposures of 10~s in low read noise mode; after every 20 of these images, two short, unsaturated co-additions of ten 0.3~s exposures in medium read noise mode were obtained to monitor the star position, the Strehl ratio, and the sky transparency. The total duration of the sequence is 1.9 hours and the total field of view rotation is 39\degr. Both before and after the main ADI sequence, a 5-position dither pattern of short unsaturated observations ($10\times0.3 s$)  was obtained to provide a sky frame for the reduction of the short exposures as well as additional image of the point-spread-function (PSF) core. 

The short, unsaturated exposures for the ADI observations were reduced as above for the $K$-band imaging. The long, saturated exposures were dark subtracted, divided by a normalized flat field image, corrected for distortion using the solution provided by the Gemini staff, and padded with zeros to increase their size to $750\times 750$ pixels and thus avoid loosing parts of the field of view during subsequent shifts and rotations. Bad pixels were corrected or masked as explained above. To compensate for possible variations in AO correction and sky transparency, each long, saturated image was intensity scaled to a common flux in an aperture of radius 4~pixels by interpolating the values measured for the short, unsaturated images obtained through the ADI sequence; these intensity corrections were  $\sim$11\% on average. The center of the stellar PSF in all saturated images was found by maximizing their cross-correlation with the closest unsaturated image in an annulus where both images are unsaturated; the images were then then shifted to position the star at the center of the frame. As a verification of the star positions determined in this way, we calculated the centroid of the planetary mass companion for all centered, de-rotated images and found a scatter of only $\sim$0.12 pixels per axis. An azimuthally symmetric intensity profile was then subtracted from each image to remove the smooth seeing halo. Next, the stellar PSF speckles were removed from each image by subtracting an optimized reference PSF image obtained using the ``locally optimized combination of images'' (LOCI) algorithm detailed in \citet{lafreniere07a}. The individual residual images were then rotated to align their FOV and their median was taken to obtain the final residual image.

\subsection{Spectroscopy}\label{sect:spectro}

The spectroscopic observations in $K$ and $H$, obtained on 2008 June 21 and 2008 August 21 and 24 with NIRI, have already been presented in \citet{lafreniere09}. The new $J$-band spectroscopic observations were obtained with the integral field spectrograph NIFS \citep{mcgregor03} on 2009 July 3. The $J$ grating was used with the $ZJ$ blocking filter, yielding a spectral resolving power of $\sim$6000. Given the lower Strehl ratio in the $J$ band and the proximity and contrast of the candidate companion relative to the primary star, the observations were obtained in ADI mode to enable a subtraction of the primary star PSF from every spectral slice of the data cube. The primary star was positioned near a corner of the $3\arcsec\times3\arcsec$ NIFS field of view and the position angle of the instrument was adjusted to ensure that the candidate companion would remain within the field of view during the ADI sequence. The individual exposure time was 240~$s$ in low read noise mode. The observations were split into two sub-sequences, of 15 and 10 exposures respectively, each one immediately followed by observation of an Ar lamp for wavelength calibration and of the A0 star HD 155379 for telluric and instrumental transmission correction. Before the first sequence and after the second sequence, two images were obtained with a dither of $\pm$15\arcsec\ to provide a sky frame. A standard NIFS flat field calibration sequence was also obtained.

\begin{figure}
\epsscale{1}
\plotone{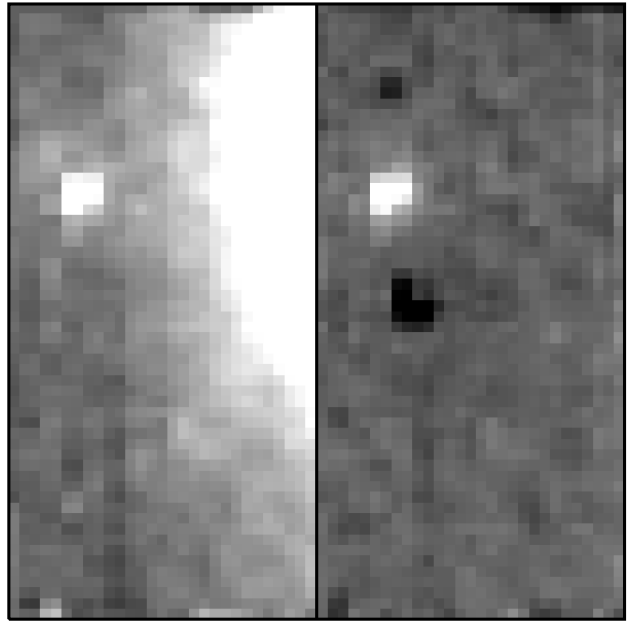}
\caption{\label{fig:nifs} Example of a median-collapsed NIFS data cube before ({\it left}) and after ({\it right}) PSF subtraction using ADI/LOCI as described in the text (\S\ref{sect:spectro}). The field of view shown is $1.5\arcsec \times 3.0\arcsec$ for each panel. The negative signals above and below the companion on the right panel are artifacts from the subtraction.}
\end{figure}

The reduction of the NIFS data, up to the reconstruction of the data cube, was made using the Gemini IRAF pipeline. The steps covered in this pipeline are sky background subtraction, flat-field and bad pixel correction, spatial and spectral calibration, and data cube reconstruction. The spatial sampling of the reconstructed data cube is 0.043\arcsec\ pixel$^{-1}$. The instrumental/telluric transmission correction and spectrum extraction, which could be done using the Gemini pipeline, were instead done using custom IDL routines as described below. First the center of the stellar PSF were registered to a common position in all slices and all cubes of the sequence; the center positions were calculated by fitting a 2D Gaussian function. The stellar PSF was then subtracted, slice by slice, using a LOCI algorithm. The `allowed' reference images included all those in which the companion had moved by at least 9 pixels (0.4\arcsec) due to field of view rotation. The optimization region was a section of annulus with inner and outer radii of 1.63\arcsec\ and 2.24\arcsec\ (14 pixel wide), respectively, excluding disks of diameter 8 pixels at each positions where the companion was present in any of the images to avoid any bias during the subtraction. Figure~\ref{fig:nifs} shows the collapsed cube before and after PSF subtraction. Following PSF subtraction, the spectrum of the companion was extracted by summing the flux in a circular aperture of diameter 5 pixels. The spectrum of the telluric standard was extracted using the same aperture, corrected for its spectral slope using a 9520~K blackbody curve, and its Paschen line at 1.28~$\mu$m was removed by dividing out a Voigt profile fit. The companion spectrum was divided by this spectrum to correct for telluric and instrumental transmission. Finally, the median of the 25 companion spectra was obtained. 

\section{Analysis and results}\label{sect:analysis}

\subsection{Verification of common proper motion}

The relative position and flux ratio of the star and companion were determined in two ways which yielded values in very good agreement. First, the centroids were determined by fitting a 2D Gaussian function to both components and their flux ratio was found using aperture photometry. Second, the relative position and relative flux were determined simultaneously by minimizing the residual noise after subtraction of one component from the other. For a given observation, these measurements were repeated for all dither positions and their average was taken. The NIRI+ALTAIR with field lens pixel scale of 21.4~mas~pixel$^{-1}$ and sky position angle from the images FITS headers were used to convert image coordinates to separations and position angles. We did not observe astrometric standards to monitor and calibrate possible changes in the pixel scale of the instrument between the epochs; such changes could occur, in principle, as a result of small changes in the shape or position of some components of the telescope, AO system or camera. Nevertheless, we can calculate upper limits on pixel scale changes using several background sources in our field of view (see \S\ref{sect:othercomp}). For this purpose we used four background stars that have been well detected at all epochs. At each epoch, we computed the mean length and position angle of all baselines between these four stars (6 baselines, lengths of 180--475 pixels). We then calculated the changes in values between the epochs for each baseline, and we finally computed, for each epoch, the average change over all baselines. The maximum pixel scale change observed was 0.14\%, corresponding to 3~mas at the separation of the companion (2.219\arcsec\ or 103.7~pixels), and the maximum change in position angle was 0.05\degr. These changes constitute upper limits as they can be affected by non-zero proper motion of the background sources. Beside these possible changes in pixel scale, our measurements are affected by larger systematic uncertainties which are estimated below. We also note that, as mentioned on the Altair webpage\footnote{\url{http://www.gemini.edu/sciops/instruments/altair/field-lens-option}}, the plate scale of NIRI when used with the field lens of Altair is not accurately calibrated. This does not affect our analysis below but we mention it as it should be considered when comparing our measurements with data from other telescopes or other instruments.

\begin{deluxetable}{cccc}
\tablewidth{0pt}
\tablecolumns{4}
\tablecaption{Astrometric measurements \label{tbl:astrom}}
\tablehead{
\colhead{Epoch} & \colhead{Band} & \colhead{$\rho$ (\arcsec)} & \colhead{P.A. (deg)}
}
\startdata
2008.3205 & $K_{\rm s}$ & $2.215\pm0.006$ & $27.75\pm0.10$ \\
2008.4596 & $H$       & $2.222\pm0.006$ & $27.76\pm0.10$ \\
2008.4596 & $J$       & $2.219\pm0.006$ & $27.76\pm0.10$ \\
2009.2621 & $K^\prime$ & $2.222\pm0.006$ & $27.65\pm0.10$ \\
2009.4972 & $K^\prime$ & $2.219\pm0.006$ & $27.74\pm0.10$
\enddata
\tablecomments{The position reported in the discovery paper was the mean of the first three rows.}
\end{deluxetable}

\begin{figure}
\epsscale{1}
\plotone{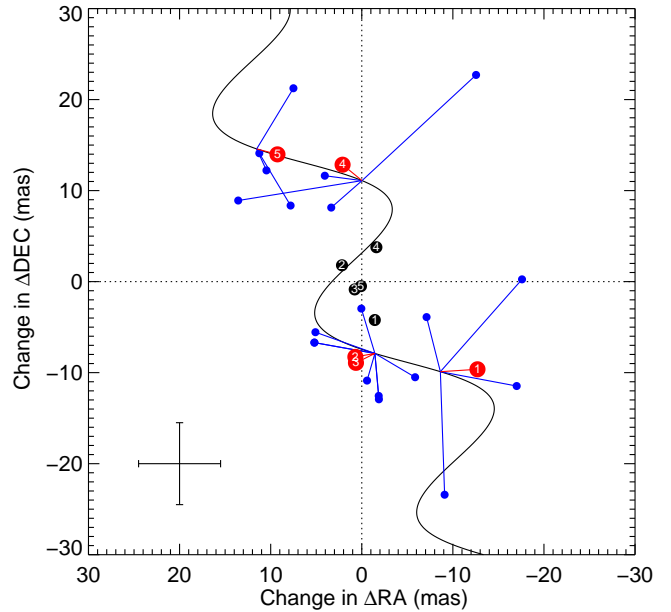}
\caption{\label{fig:pm}  Measured changes in offset between the companion candidate and the primary star (black filled circles) over the different epochs compared with the expected changes in offset for a distant stationary background source (solid black curve), as calculated from the proper motion and distance of the primary star. The measured changes in offset between four background stars and the primary star are also shown with small blue circles. The big red filled circles show the mean changes of the four background stars at each epoch. The blue and red circles are are connected with the background model curve at their corresponding epoch. The numbers inside the circles indicate the epoch according to the order of Table~\ref{tbl:astrom}. The error symbol at the lower left shows the uncertainty for individual data points. For all sources, the mean offset change over all epochs has been subtracted.}
\end{figure}

Table~\ref{tbl:astrom} provides a list of our astrometric measurements and figure~\ref{fig:pm} shows the changes in offset between the companion and primary over all epochs compared with the expected changes for a distant background star, based on the proper motion and estimated distance of the primary; the figure also shows the same measurements for four background stars. For the proper motion of the primary, we used the value given in the UCAC3 catalog \citep{zacharias09}: $-11.2\pm1.5$~mas~yr$^{-1}$ in RA and $-21.9\pm1.5$~mas~yr$^{-1}$ in DEC. A few other values, differing by less than 6~mas~yr$^{-1}$ per axis, are also available in the literature but have larger uncertainties; using those other values would not affect our conclusions. For the parallax motion we assumed the mean distance and spread of the Upper~Sco association given in \citet{preibisch02}, namely $145\pm20$~pc. As visible on the figure, the background star measurements are in good agreement with the background source model, although there are large systematic errors for the individual data points. These systematic errors are likely due in part to residual geometric distortion. The average absolute difference between the measurements and the model for the background stars is 4.5~mas in both RA and DEC, or  6~mas in separation and 0.1\degr\ in position angle; we adopt these values as our overall astrometric errors. Figure~\ref{fig:pm} also clearly shows that, within our uncertainties, the companion's position relative to the primary did not change over time; conversely, the measurements for the companion are inconsistent with the background model. From the earliest to the latest epoch, the measurements are $\sim6\sigma$ off from the background source hypothesis. The $\chi^2$ values for the common proper motion or background source hypotheses are 2.3 and 33, respectively. A simple linear fit of the astrometry measurements with time yields changes of $2.0\pm5.6$~mas~yr$^{-1}$ for the separation and $-0.05\pm0.09$~deg~yr$^{-1}$ for the position angle (or $-0.9\pm4.0$~mas~yr$^{-1}$ in RA and $2.7\pm5.2$~mas~yr$^{-1}$ in DEC), providing limits on the differential motion of the companion. Our precision is insufficient, however, to constrain the orbital motion of the companion, expected to be $\sim$2.2~mas~yr$^{-1}$ assuming that the planet semi-major axis is close to its projected separation.

Our new measurements clearly confirm the common proper motion of the two objects, but as pointed out in Paper~I this does not in itself prove physical association as the velocity dispersion in the Upper~Sco association is small. We previously presented an analysis indicating that, conservatively, the probability that a planetary mass object in Upper~Sco would fall within 2.5\arcsec\ of any of the stellar members we observed is only 0.002. We can now factor the common proper motion constraint into the analysis. Using the proper motion dispersion of $\sim$8~mas~yr$^{-1}$ in RA and DEC reported by \citep{bouy09} for the low-mass members of Upper~Sco, the probability that two unrelated Upper~Sco objects would share a common proper motion within 4-5~mas~yr$^{-1}$ per axis, as we have observed, is 0.13. Thus taken together, the proximity and common proper motion of the primary and companion indicate a probability of $2.6\times10^{-4}$ of chance alignment. So given this very small probability, despite the lack of orbital motion detection, we treat the candidate companion as a truly bound companion for the remainder of this paper. 

\subsection{Properties of the companion}

\begin{figure}
\epsscale{1}
\plotone{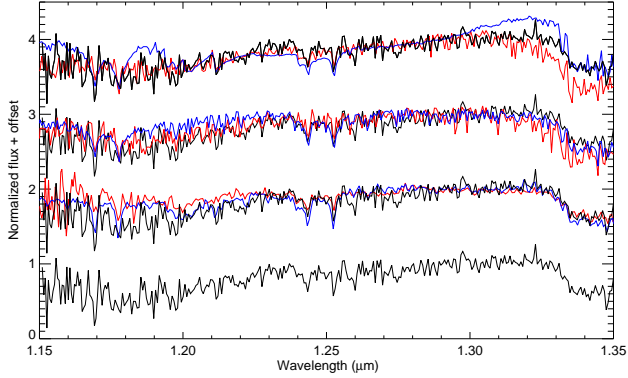}
\caption{\label{fig:jspec}  $J$-band spectrum of \companion\ (black) along with various comparison spectra (red and blue). From the top, the first spectrum is compared with synthetic spectra from the DUSTY models with $T_{\rm eff}=1800$~K and $\log{g}=3.5$ (red) or $\log{g}=6.0$ (blue); the second is compared with synthetic spectra from the DRIFT PHOENIX models with $T_{\rm eff}=1800$~K and $\log{g}=3.5$ (red) or $\log{g}=6.0$ (blue); and the third is compared with the spectrum of the L0 brown dwarf USco~J160606-233513 (red), which is a known member of the Upper~Sco association \citep{lodieu08}, and the spectrum of the field L3 dwarf 2MASS~J1506544+132106 (blue, \citealp{cushing05}). The spectra are normalized at $\sim$1.29~$\mu$m and binned to a resolving power of 2000.}
\end{figure}

%spectroscopy
The new $J$-band spectrum of the companion is shown in Fig.~\ref{fig:jspec} along with the spectrum of a field L3 dwarf, the spectrum of a young L0 brown dwarf member of Upper~Sco, and model spectra from the DUSTY \citep{chabrier00} and DRIFT PHOENIX \citep{witte09,helling08} atmosphere models for low and high surface gravity. The new $J$-band spectrum shows typical features of late-M or early-L dwarfs, namely important absorption by H$_2$O beyond $\sim$1.33~$\mu$m, the K~I doublet at 1.24-1.25~$\mu$m, and absorption by FeH at 1.24~$\mu$m; the S/N is too small to identify other individual features but the general shape of the continuum is qualitatively as expected through the band. As seen from the reasonable agreement with the model spectrum of $T_{\rm eff}=1800$~K, the new spectrum is consistent with the previous estimate of the effective temperature of the companion. 

\begin{figure}
\epsscale{1}
\plotone{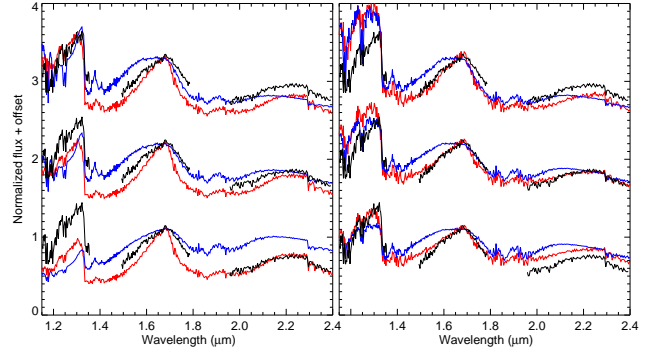}
\caption{\label{fig:jhkspec} Merged $JHK$ spectrum of \companion\ (black) compared with various synthetic spectra from the DUSTY (left) and DRIFT PHOENIX (right) atmosphere models. From top to bottom, the comparison spectra have $T_{\rm eff}$ of 2000~K, 1800~K and 1700~K. The red curves are for $\log{g}=4.0$ and the blue curves are for $\log{g}=6.0$. The spectra are normalized at 1.67~$\mu$m and binned to a resolving power of 500. Uncertainties in the colors of the companion translate into $\sim$15\% uncertainty in the relative scaling of the companion spectrum between the bands.}
\end{figure}

Figure~\ref{fig:jhkspec} shows the merged $JHK$ spectrum of \companion\ along with synthetic spectra from the DUSTY and DRIFT PHOENIX atmosphere models for different temperatures and surface gravities. From this figure, it is clear that the companion has low surface gravity: the slope of the continuum through the $H$ and $K$ bands are much better fit by low gravity spectra for both sets of models. The DRIFT PHOENIX models provide a better fit of the overall spectral shape. In particular, the red side of the $H$ band, which is poorly fit by the low-gravity DUSTY models owing to uncertainties in the model opacities used, is well fit the by the DRIFT PHOENIX models. For the latter models, the steepness of the slope on either side of the $H$ band even provides some constraint on the effective temperature, favoring 1700--1800~K. Although less apparent from the figure, the $K$ band is also better reproduced by the DRIFT PHOENIX models, the DUSTY models spectra falling off slightly too rapidly at the blue side. The shape of the $K$ band spectrum, according to the DRIFT PHOENIX models, also favors a temperature of 1800~K.

\begin{figure}
\epsscale{1}
\plotone{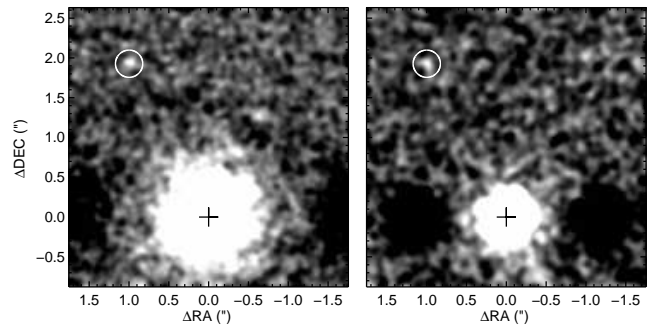}
\caption{\label{fig:longlbd} Altair/NIRI images of \primary\ in the 3.05~$\mu$m narrow band filter ({\it left}) and in the $L^\prime$ filter ({\it right}). The white circle in each panel marks the expected position of the companion. The images have been convolved by a Gaussian kernel to improve the S/N of the companion. The negative regions left and right of the primary star in each panel are artifacts from the sky subtraction.}
\end{figure}

%long lambda photometry
The companion was detected in both the NB3.05 and $L^\prime$ images at the level of $\sim$5$\sigma$ and $\sim$3$\sigma$, respectively; see Fig.~\ref{fig:longlbd}. The flux of the companion and primary were determined from photometry in an aperture of diameter 6 pixels. The uncertainty on the companion flux was estimated from the dispersion of the noise in 21 such apertures spread azimuthally around a circle of radius 2.2\arcsec\ centered on the primary star. We obtained a contrast of $6.8\pm0.2$~mag in NB3.05 and $6.1\pm0.3$~mag in $L^\prime$. 

Unfortunately, we cannot derive the companion's flux in a straightforward manner as we did not observe photometric standard stars to calibrate the photometry, and no measurements of the primary flux in the same filters exist in the literature. However there exist calibrated flux measurements of the primary in several infrared bands that closely bracket the wavelength range of our observations: 2MASS $J$, $H$ and $K_{\rm s}$ \citep{2mass}, Spitzer IRAC 4.5~$\mu$m and 8.0~$\mu$m, and Spitzer IRS peak-up imaging at 16~$\mu$m \citep{carpenter06}. We used these measurements, together with model spectra, to estimate the flux of the primary in the NB3.05 and $L^\prime$ filters. More specifically, we used synthetic spectra from the NextGen models \citep{hauschildt99} to compute synthetic fluxes in all the bands listed above, we then adjusted those synthetic fluxes to the measured values, and finally computed the synthetic average fluxes of the primary star in the NB3.05 and $L^\prime$ bands. For the 2MASS bands, the synthetic average flux densities were computed using the relative spectral response curves given on the 2MASS project webpage\footnote{\url{http://www.ipac.caltech.edu/2mass/releases/allsky/doc/sec6\_4a.html}} and the flux zero points determined by \citet{rieke08}; for the Spitzer IRAC and IRS bands, the spectral response curves and zero points were taken from the respective instruments webpages\footnote{see \url{http://ssc.spitzer.caltech.edu/irac/calibrationfiles} \\ and \url{http://ssc.spitzer.caltech.edu/irs/calibrationfiles/}}. Various synthetic spectra were adjusted to the photometric measurements by minimizing the reduced $\chi^2$. The spectrum yielding the best fit, $T_{\rm eff}=4000$~K and $\log{g}=4.0$ was used in combination with the spectral response curves of the NIRI NB3.05 and $L^\prime$ filters, taken from the NIRI website\footnote{\url{http://www.gemini.edu/sciops/instruments/niri/}}, to compute the primary star flux density in those two bands. We obtained values of $1.71\times10^{-14}$~W~m$^{-2}$~$\mu$m$^{-1}$ for NB3.05 and $3.36\times10^{-14}$~W~m$^{-2}$~$\mu$m$^{-1}$ for $L^\prime$. Then using the observed contrast for the companion we obtained fluxes of $(6.4\pm1.3)\times10^{-17}$~W~m$^{-2}$~$\mu$m$^{-1}$ for NB3.05 and $(6.2\pm1.8)\times10^{-17}$~W~m$^{-2}$~$\mu$m$^{-1}$ for $L^\prime$.

To verify the accuracy of the above procedure in estimating the primary star flux, we repeated it by ignoring one band ($K_{\rm s}$ or IRAC~4.5~$\mu$m) when adjusting the synthetic fluxes to the observations and then retrieving the synthetic flux in that band from the adjusted spectrum. For both $K_{\rm s}$ and IRAC~4.5~$\mu$m, the flux retrieved was within 3\% from the measured value. Thus the error on the primary star flux resulting from this procedure is a small contribution to the error on the companion flux.

\begin{figure}
\epsscale{1}
\plotone{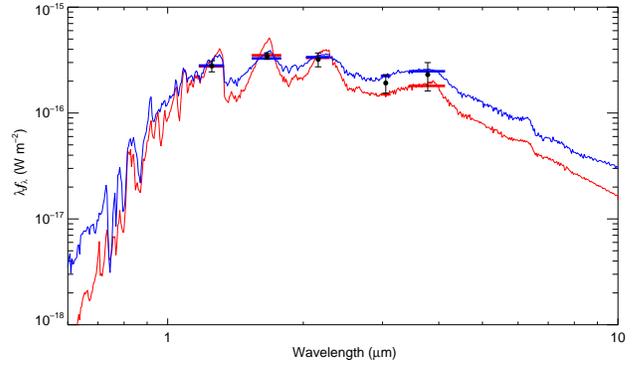}
\caption{\label{fig:sed}  Photometric measurements of the companion (black points with error bars) compared with the DUSTY model spectrum (red curve) for $T_{\rm eff}=1800$~K and $\log{g}=4.0$ and the DRIFT PHOENIX model spectrum (blue curve) for $T_{\rm eff}=1700$~K and $\log{g}=4.0$. The short horizontal lines mark the corresponding model synthetic fluxes in each band; the length of the lines marks the extent of each photometric band.}
\end{figure}

\begin{figure}
\epsscale{1}
\plotone{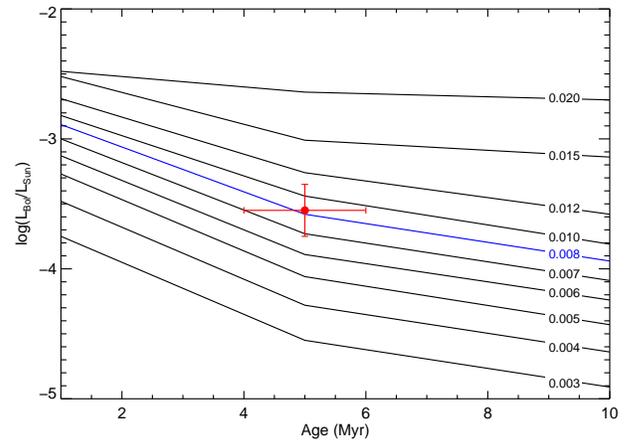}
\caption{\label{fig:lbol} Calculated bolometric luminosity of the companion (point) compared with the DUSTY model evolutionary tracks for different masses (curves). }
\end{figure}

%mass of companion
In \citet{lafreniere08b} we derived a mass estimate for the companion primarily from our estimate of its effective temperature, and commented that the value found was consistent with the observed $K_{\rm s}$ brightness. Here we adopt a different approach based on the bolometric luminosity of the companion inferred from a fit of model spectra to all photometric measurements available. The synthetic spectra were adjusted to the photometry in the same way as done above for the primary. After adjustment to the data points, the model spectra were integrated over all wavelengths to obtain the total irradiance, which was then converted to bolometric luminosity using the primary star distance estimate of $145\pm20$~pc. We repeated this procedure using both the DUSTY and DRIFT PHOENIX model spectra for a range of $T_{\rm eff}$ from 1600~K to 2000~K and $\log{g}$ from 3.5 to 6.0. For both sets of models, the spectra yielding the best agreement with the data points have $T_{\rm eff}=$1700--1800~K and $\log{g}=$3.5--4.5; representative spectra are shown in Fig.~\ref{fig:sed}. Despite significant differences in the model spectra over the range explored, the range of bolometric luminosities obtained ($\pm0.08$~dex) is rather small, as noted by \citet{marois08} who used the same procedure for the HR~8799 planets. Then accounting for the uncertainty in distance, we obtain a bolometric luminosity of $\log{(L/L_\odot)}=-3.55\pm0.2$. Compared to the prediction of the DUSTY evolution models (see Fig.~\ref{fig:lbol}), this luminosity indicates a mass of $0.008_{-0.002}^{+0.003}$~\msun, in perfect agreement with the value reported in \citet{lafreniere08b}. For this mass, the dusty evolution models predict a $T_{\rm eff}$ of 1800~K, fully consistent with the best spectral fit. Thus, notwithstanding uncertainties in evolution models, the mass of \companion\ is in the planetary regime.

We also used the same procedure, with NextGen model spectra of $T_{\rm eff}=$3800-4000~K (K7-M0) and $\log{g}=$3.5-4.5, to estimate the bolometric luminosity of the primary star. We obtained a value of $\log{(L/L_\odot)}=-0.37\pm0.15$, in good agreement with the prediction (-0.36) of the evolution models of \citet{baraffe98} with $\alpha_{\rm mix}=1.9$ for an age of 5~Myr and $T_{\rm eff}$ of 4000~K. The ratio of the companion bolometric luminosity to that of the companion, which is independent of our distance uncertainty, is thus $\log{(L_{\rm comp}/L_{\rm prim})}=-3.18\pm0.09$.

\subsection{Search for other companions}\label{sect:othercomp}

\begin{figure}
\epsscale{1}
\plotone{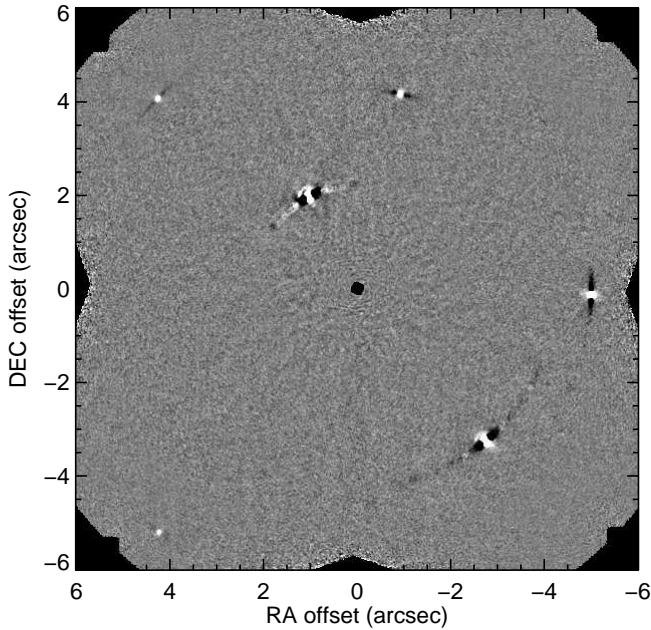}
\caption{\label{fig:adi} Residual ADI image normalized by the noise radial profile. The display is from $-5\sigma$ to $+5\sigma$. The azimuthal positive/negative trails around the visible sources are artifact from the PSF subtraction algorithm. }
\end{figure}

The residual ADI image is shown in Fig.~\ref{fig:adi}. In addition to the companion, this image shows 5 other point sources that are clearly detected. These sources are all further away from the primary and fainter than the companion. All of these five sources were detected at least twice in our multiple epoch normal imaging and the astrometric measurements were sufficient to determine that they are background stars. For reference, the position and contrast of these sources are given in Table~\ref{tbl:background}. Visual inspection of the image did not reveal any other new, fainter point source.

\begin{deluxetable}{ccc}
\tablewidth{0pt}
\tablecolumns{3}
\tablecaption{Background sources detected \label{tbl:background}}
\tablehead{
\colhead{$\rho$ (\arcsec)\tablenotemark{a}} & \colhead{P.A. (deg)\tablenotemark{a}} & \colhead{$\Delta K^\prime$ (mag)}
}
\startdata
$4.240\pm0.006$ & $220.07\pm0.10$ & $8.40\pm0.10$ \\
$4.251\pm0.006$ & $347.57\pm0.10$ & $10.65\pm0.10$ \\
$4.687\pm0.006$ & $268.51\pm0.10$ & $9.50\pm0.10$ \\
$5.878\pm0.006$ & $46.40\pm0.10$ & $10.9\pm0.2$ \\
$6.706\pm0.006$ & $140.84\pm0.10$ & $11.4\pm0.2$
\enddata
\tablenotetext{a}{At epoch 2009.2621.}
\end{deluxetable}

\begin{figure}
\epsscale{1}
\plotone{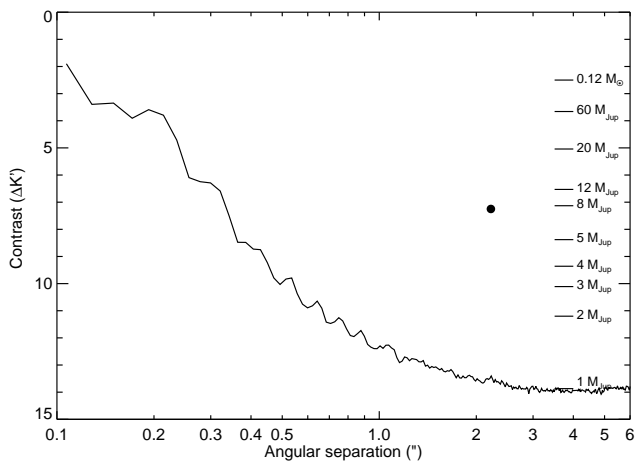}
\caption{\label{fig:adilim}  Detection limits ($5\sigma$) achieved by the ADI observations. The expected contrast for various planetary masses are shown toward the right of the figure. The filled circle marks the companion, \companion.}
\end{figure}

Detection limits in difference of magnitude as a function angular separation were computed from the standard deviation of the pixel values in an annulus of width equal to one PSF FWHM; the photometry was properly calibrated using the procedure described in \S3.3 of \citet{lafreniere07}. The result is shown in Fig.~\ref{fig:adilim}.  The evolution models of \citet{baraffe03}, \citet{chabrier00} and \citet{baraffe98} were used to translate these detection limits into planetary masses, assuming an age of 5~Myr. Our results allow us to rule out the presence of planets more massive than 1~\mjup, 2~\mjup\ and 8~\mjup\ at projected separations larger than 3\arcsec\ ($\sim$440~AU), 0.7\arcsec\ ($\sim$100~AU) and 0.35\arcsec\ ($\sim$50~AU), respectively. These limits indicate that if the planetary companion migrated outward (or was ejected) to its current location through planet-planet scattering, then the perturbing body, likely more massive, would have to be located within 50~AU.

\section{Concluding remarks}\label{sect:conclusion}

The planetary mass companion \companion\ is the least massive known to date with an orbital separation of a few hundreds of AU. As discussed in Paper~I, the existence of such a low mass companion orbiting so far away from its star poses a great challenge for star and planet formation models. Indeed, all main modes of formation --core accretion, gravitational instability or binary-like formation-- and migration --disk interaction or planet-planet interaction-- face some obstacles when trying to account for such a companion. A lot of work has been done over the past two years in an attempt to explain the handful of companions at moderate-to-large separations recently found by direct imaging, the general outcome being that, although it remains challenging, given the right initial conditions, the above mechanisms can indeed succeed in producing distant gas giants. For example, gravitational scattering between multiple planets is a viable explanation for the planet we have found at $\sim$330~AU \citep[e.g.][]{scharf09, veras09}, but of course this would require that several massive planets have successfully formed at smaller orbital separations ($<$30--100~AU). If this is the case, then \companion\ should have a highly eccentric orbit and would likely eventually be ejected from the system. Alternatively, some authors have found that gravitational instability may also be a viable explanation, provided that the disk of this system was sufficiently massive, extended and had a relatively low opacity \citep[e.g.][]{meru10, boley10}; although admittedly, the regime where gravitational instability is effective remains a matter of debate \citep[e.g.][]{kratter10, stamatellos08, rafikov07, boss06}.

More follow-up observations of this system over the coming years should help establish which, if any, of these mechanisms can best explain the origin of this planet. In particular, multi-epoch high-resolution spectroscopy of the primary and additional high-contrast imaging with improved sensitivity below 0.35\arcsec\ would allow to search for even closer-in companions; this would guide migration and ejection scenarios involving dynamical interaction between multiple planets. Also, higher S/N and/or higher resolution spectroscopy of the companion would allow to constrain the differential radial velocity of the primary and companion. This would constrain the orbital velocity of the companion --and its true physical bound-- or its possible recent ejection. These data could also be used to constrain the metallicity of the planet, which could provide some clues to its formation mechanism. Better spectroscopy would also enable a much more detailed comparison with model spectra. Along the same line, photometry over an extended spectral range would be useful to verify model spectra. A parallax measurement for the primary star would remove a big part of the uncertainty on the companion luminosity, which would then lead to a better mass estimate and/or a more constraining verification of evolution models. Finally, continued astrometry monitoring should begin to constrain the orbital motion of the companion within a few years, given the expected differential motion of $\sim$2.2~mas~yr$^{-1}$.

\acknowledgments

We thank the Gemini staff for help and support with the observations reported in this paper. We also thank Shami Chatterjee for preparation and analysis of the VLA observations. We are grateful to Christiane Helling for making the DRIFT PHOENIX spectra available to us. Finally, we would like to thank our referee for excellent suggestions that improved the quality of this paper.

\end{document}